\documentclass[twocolumn,prb]{revtex4}
\usepackage{amsmath}
\usepackage{graphicx}
\usepackage{epsfig}
\begin{document}

\title{Dynamics of interacting Brownian particles: a diagrammatic formulation}
\author{Grzegorz Szamel}
\affiliation{Department of Chemistry, 
Colorado State University, Fort Collins, CO 80525}

\date{\today}

\begin{abstract}
We present a diagrammatic formulation of a theory for the time dependence of 
density fluctuations in equilibrium systems of interacting Brownian particles. 
To facilitate derivation of the diagrammatic expansion we introduce a basis that  
consists of orthogonalized many-particle density fluctuations. We obtain an exact 
hierarchy of equations of motion for time-dependent correlations of
orthogonalized density fluctuations. 
To simplify this hierarchy we neglect contributions to the vertices 
from higher-order cluster expansion terms. An iterative solution
of the resulting equations can be represented by diagrams with three and 
four-leg vertices. We analyze the structure of the diagrammatic series 
for the time-dependent density correlation function 
and obtain a diagrammatic interpretation
of reducible and irreducible memory functions. 
The one-loop self-consistent approximation for the latter 
function coincides with mode-coupling approximation for Brownian systems 
that was derived previously using a projection operator approach.
\end{abstract}
\maketitle

\section{Introduction}

There has been a lot of interest in recent years in the dynamics of interacting
Brownian particles \cite{th,review}. 
The reason for this interest is twofold. First, experiments have provided a wealth 
of information about the motion of individual colloidal particles \cite{expts}. 
A system of interacting Brownian particles is the simplest model of a colloidal 
suspension. Second, interacting Brownian particles constitute 
the simplest model system on which one can test techniques and approximations
of non-equilibrium statistical
mechanics. It is a simpler model system than a simple fluid for both fundamental 
reasons (irreversibility is built in) and 
technical reasons (particles have fewer degrees 
of freedom due to the overdamped character of their motion). 

In this paper we present a diagrammatic approach to the description of 
time-dependent density fluctuations in equilibrium
systems of interacting Brownian particles (here an equilibrium
system means a stable or a metastable (\textit{e.g.} supercooled) equilibrium system). 
The original inspiration for this work was 
a series of three papers \cite{H1,H2,H3} by
Hans Andersen in which a general framework of a diagrammatic approach to 
the dynamics of fluctuations in equilibrium simple fluids was presented. 
An important feature of Andersen's approach was adoption of a specific set of basis
functions, termed Boley \cite{Boley} basis. 
As lucidly explained in Ref. \cite{H2}, one of the advantages of this set of basis 
functions is an enormous simplification of the initial condition for the 
whole hierarchy of equations for time-dependent correlation functions \cite{Opp}.
Additional motivation for our work comes from renewed 
interest  \cite{field1,field2,field3,field4} 
in developing field theories for systems of strongly 
interacting particles and in using these theories to generate approximate 
self-consistent approaches to the dynamics of these systems.
Such field theories usually lead to diagrammatic series for so-called 
response and correlation functions. Our work might be the first
step in a reverse procedure: constructing a field theory from a diagrammatic
approach. Also, our approach provides a very simple derivation of the mode-coupling
theory \cite{Goetze} that has been extensively used to describe colloidal systems. 
This theory has been previously derived using a projection operator 
approach \cite{SL}. 
More recent, field-theoretical derivations have 
either been found unsatisfactory \cite{KM} or are quite involved \cite{field2,field3}. 
Finally, new techniques have recently been developed for strongly
correlated many-body quantum systems that allow one to numerically 
integrate \cite{diagramMC} and approximately analyze \cite{diagramapproxsum} 
whole diagrammatic series. It is hoped that these methods
could be adopted to classical many-body systems and, in particular,
that they could be used to evaluate diagrammatic series presented in this paper.   

Our diagrammatic approach to the dynamics of equilibrium systems of interacting 
Brownian particles is similar to that developed by Andersen \cite{H1,H2,H3}
for simple fluids. In spite of the fact that a Brownian 
system is simpler  than a simple fluid, in the present problem it is 
advantageous to introduce two different sets of basis functions. 
As a consequence, a general structure that leads to the
emergence of so-called irreducible memory function appears naturally 
in the diagrammatic expansion. 
Our approach uses one important aspect of Andersen's work: in Ref. \cite{H1}
the existence of a basis of orthogonalized many-particle phase-space density 
fluctuations was established. We use a consequence
of this result: we assume the existence of a basis consisting 
of orthogonalized many-particle
density fluctuations in the Fourier space. We also assume the existence of a second, 
closely-related orthogonalized basis of many-particle self-density fluctuations 
in the Fourier space.
The latter basis was used previously in the description of self-diffusion
in Brownian systems \cite{SLeg,Gauss}. Our main, formal result is a hierarchy 
of equations for time-dependent 
correlations of the orthogonalized many-particle densities. An important
feature of this hierarchy is that all the interactions are renormalized: they are 
expressed in terms of equilibrium correlation functions. To simplify the 
structure of the hierarchy, we neglect the contributions to the terms describing
inter-particle interactions (\textit{i.e.} vertices) coming from  
higher-order cluster expansion terms. 
An iterative solution of the simplified hierarchy can be interpreted in
terms of diagrams. After some simplifications we obtain an expansion in
terms of diagrams consisting of lines corresponding to
free diffusion and three-, and four-leg vertices. 
We analyze the structure
of the diagrammatic expression for the density correlation function and 
show that so-called irreducible memory function appears in a very natural way.
Finally, we present a diagrammatic derivation of the standard, G\"{o}tze-like 
\cite{Goetze,SL} mode-coupling approximation.

The paper is organized as follows. In Sec. \ref{basis}, we introduce two sets
of basis functions. In Sec. \ref{eofm}, we derive exact, formal equations of
motion for time-dependent correlations of orthogonalized many-particle densities
and in Sec. \ref{eofmapprox}, we simplify these equations by neglecting
contributions to the vertices from higher order cluster expansion terms. 
Sec. \ref{diagram} is devoted to the derivation of  diagrammatic
representation: first, the 
approximate equations of motion are re-written as integral equations; then, the
iterative solution of the latter equations is interpreted in terms
of labeled diagrams; finally, a series expansion in terms of labeled diagrams 
is rewritten in terms of unlabeled diagrams. In Sec. \ref{mf}, we analyze
the series expansion and present diagrammatic
expressions for so-called memory and irreducible memory functions. In Sec. \ref{mct},
we show that a self-consistent one-loop approximation for the 
irreducible memory function is equivalent to the mode-coupling approximation.
We close in Sec. \ref{discussion} with a discussion of our results, 
a comparison with other approaches, and an outline of future research.

\section{Basis functions: orthogonalized many-particle densities}\label{basis}

We consider a system of $N$ interacting Brownian particles in volume $V$.
The average density is $n=N/V$. The brackets $\left< ... \right>$ 
indicate a canonical ensemble average at temperature $T$. In Secs. \ref{basis} 
and \ref{eofm}, we consider a large but finite system and in Sec. \ref{eofmapprox}, 
we take the thermodynamic limit, $N\to\infty, V\to\infty, N/V=n=const.$

We start by introducing a set of Fourier transforms of many-particle densities,
\begin{equation}\label{nludef}
N_k(\mathbf{k}_1,...,\mathbf{k}_k) = 
\sum_{i_1\neq ...\neq i_k=1}^{N} 
e^{-i\mathbf{k}_1\cdot\mathbf{r}_{i_1}-  ... 
- i\mathbf{k}_k\cdot\mathbf{r}_{i_k}}.
\end{equation}
Here $k> 0$ and $\mathbf{r}_i$, $i=1, ..., N$ denote positions of the particles.
For simplicity, we will henceforth use term many-particle densities for the
Fourier transforms of these densities. Also, we will sometimes use abbreviated
notation. Hence $N_k(\mathbf{k}_1,\mathbf{k}_2,...,\mathbf{k}_k)$ may be
written as $N_k(1,2,...,k)$ or even as $N_k$. Also, sum over wavevectors,
$\sum_{\mathbf{k}_1,...,\mathbf{k}_k}$ may be written as $\sum_{1,...,k}$. 
It should be noted that 
densities $N_k$ are symmetric functions of their arguments. 

Following Andersen \cite{H1}, we introduce orthogonalized many-particle densities
using the language of a Hilbert space. The densities are mapped onto vectors, 
\begin{equation}\label{vectors}
N_k(\mathbf{k}_1,...,\mathbf{k}_k) \leftrightarrow 
\left|N_k(\mathbf{k}_1,...,\mathbf{k}_k)\right>,
\end{equation} 
and the scalar product is defined as
\begin{equation}\label{scalar}
\left<N_k | N_l \right> = \left<N_k N_l^* \right>,
\end{equation}
where the asterisk denotes complex conjugation.

To define a set of vectors corresponding to the 
orthogonalized many-particle densities we start from the
0-particle density,
\begin{equation}\label{n0def}
\left|n_0\right> \leftrightarrow  n_0 \equiv N.
\end{equation}
Next, we introduce a projection operator $\mathcal{P}_0$ 
onto a subspace spanned by $\left|n_0\right>$, and 
define $ \left|n_1\right>$ as the part of $ \left|N_1\right>$
that is orthogonal to $\left|n_0\right>$,
\begin{equation}\label{n1def}
\left. |n_1\right> \equiv \left( 1 - \mathcal{P}_0 \right) \left. |N_1\right>.
\end{equation}
Having introduced $\left|n_1\right>$ we can define a projection operator 
$\mathcal{P}_1$ onto a subspace spanned by it. This allows us to define 
$\left|n_2\right>$ as the part of $ \left|N_2\right>$
that is orthogonal to $\left|n_0\right>$ and $\left|n_1\right>$,
\begin{equation}\label{n2def}
\left|n_2\right> \equiv \left( 1 - \mathcal{P}_0 - \mathcal{P}_1 \right) 
\left|N_2\right>.
\end{equation}
Higher-order orthogonalized many-particle densities can be introduced by continuing
this recursive procedure. The 
orthogonalized densities are symmetric functions of their arguments.
The set of the
orthogonalized densities constitutes the Boley \cite{Boley} 
basis for the present problem.

It should be emphasized that the orthogonalization procedure described above 
implicitly assumes 
the existence of the projection operators \cite{H1}. The simplest, trivial example
is that of $\mathcal{P}_1$. We can write $\mathcal{P}_1$ as 
\begin{equation}\label{P1exis1}
\mathcal{P}_1 = \sum_{1,2}\left|n_1(1)\right>K_1(1;2)\left<n_1(2)\right|
\end{equation}
where $K_1$ is the inverse of the $F_1(1;2)=\left<n_1(1)|n_1(2)\right> 
\equiv \left<n_1(1)n_1^*(2)\right>$,
\begin{equation}\label{P1exis2}
\sum_{\mathbf{k}_3} \left<n_1(\mathbf{k}_1)n_1^*(\mathbf{k}_3)\right> 
K_1(\mathbf{k}_3,\mathbf{k}_2) = \delta_{\mathbf{k}_1,\mathbf{k}_2}.
\end{equation}
One notes immediately that $\left<n_1(\mathbf{k}_1)n_1^*(\mathbf{k}_3)\right> = 
NS(k_1)\delta_{\mathbf{k}_1,\mathbf{k}_3}$, where $S(k)$ is the static structure factor, 
and thus 
$K_1(\mathbf{k}_3;\mathbf{k}_2) = (1/(NS(k_2))) \delta_{\mathbf{k}_3,\mathbf{k}_2}$.

In general, we can formally write 
\begin{eqnarray}\label{Pkexis1}
\mathcal{P}_k &=& \frac{1}{(k!)^2} \sum_{1,...,k,1',...,k'}
\left|n_k(1,..,k)\right>K_k(1,..,k;1',...,k')\nonumber \\ && 
\times \left<n_k(1',...,k')\right|.
\end{eqnarray}
Here $K_k$ is the inverse of $F_k(1,...,k;1',...,k')$,
\begin{eqnarray}\label{Fkdef}
F_k(1,...,k;1',...,k') &=& \left<n_l(1,...,k) | n_k(1',...,k')\right> \nonumber \\
&\equiv &\left<n_k(1,...,k)n_k^*(1',...,k')\right>,\;\;\;
\end{eqnarray}
\begin{eqnarray}\label{Pkexis2}
\lefteqn{\frac{1}{k!} \sum_{1'',...,k''}  \left<n_k(1,...,k)n_k^*(1'',...,k'')\right>}
\nonumber \\ && \times
K_l(1'',...,k'';1',...,k') = I_l(1,...,k;1',...,k').\;\;\;
\end{eqnarray}
In Eq. (\ref{Pkexis2}) $I_k(1,...,k;1',...,k')$ is an identity defined as
\begin{equation}\label{Ikdef}
I_l(1,...,k;1',...,k') = \sum_{\wp(1',...,k')} \delta_{1,1'}...\delta_{k,k'},
\end{equation}
where $\wp(1',...,k')$ denotes a permutation of the arguments $1',...,k'$, and the sum
is over $k!$ distinct permutations.

The question of the existence of functions $K_k$ 
is related to the question of the existence of similar functions that was  
discussed and answered affirmatively 
in Sec. 3 of Ref. \cite{H1} (a careful reader will have by now
noticed that we partially follow notation introduced in that paper). 
The only, minor difference is that the functions
considered in this work are Fourier transforms of the many-particle densities in 
position space whereas the functions considered in Ref. \cite{H1} are many-particle
densities in phase-space. 

It will become clear in the next section that in addition to the set of densities 
$n_k$, it is advantageous to introduce another set of 
orthogonalized densities. This set of densities was implicitly used in 
investigations of self-diffusion in Brownian systems \cite{SLeg,Gauss}.

We start with the self-density,
\begin{equation}\label{ns1udef}
N^s_1(\mathbf{k}_1) = e^{-i\mathbf{k}_1\cdot\mathbf{r}_1}.
\end{equation}
$N_1^s$ depends on the particle number 1; note that 
there is nothing special about selecting
this particular particle and any other particle can be used in its place.

Next, we define analogous many-particle self-densities,
\begin{equation}\label{nsludef}
N^s_k(\mathbf{k}_1,...,\mathbf{k}_k) = 
\sum_{i_1\neq...\neq i_{k-1}=2}^{N} 
e^{-i\mathbf{k}_1\cdot\mathbf{r}_1-i\mathbf{k}_2\cdot\mathbf{r}_{i_1} - ... 
- i\mathbf{k}_k\cdot\mathbf{r}_{i_{k-1}}},
\end{equation}
and associated vectors in the Hilbert space,
\begin{equation}\label{svectors}
N_k^s(\mathbf{k}_1,...,\mathbf{k}_k) \leftrightarrow 
\left|N_k^s(\mathbf{k}_1,...,\mathbf{k}_k)\right>,
\end{equation} 
It should be noted that self-densities $N^s_k(1,2,...,k)$ are symmetric functions
of their $k-1$ last arguments.

Finally, we perform a recursive orthogonalization.
To make this procedure similar to that used for many-particle densities 
we start with the 0-particle self-density,
\begin{equation}\label{n0sdef}
\left|n_0^s\right> \leftrightarrow  n_0^s \equiv 1,
\end{equation}
and then we define the 1-particle self-density,
\begin{equation}\label{n1sdef}
\left. |n_1^s\right> \equiv \left( 1 - \mathcal{P}_0^s \right) \left. |N_1^s\right>,
\end{equation}
where $\mathcal{P}_0^s$ is a projection operator on a subspace spanned
by $\left|n_0^s\right>$.
Next, we introduce a projection operator $\mathcal{P}_1^s$ 
onto a subspace spanned by $\left|n_1^s\right>$ and 
we define $ \left|n_2^s\right>$ as the part of $ \left|N_2^s\right>$
that is orthogonal to $\left|n_0^s\right>$ and $\left|n_1^s\right>$, 
\begin{equation}\label{n2sdef}
\left|n_2^s\right> \equiv 
\left( 1 - \mathcal{P}_0^s - \mathcal{P}_1^s \right) \left. |N_2^s\right>.
\end{equation}
Again, higher-order orthogonalized self-densities can be introduced
by continuing this procedure.

As before, the orthogonalization procedure relies upon the existence of projection
operators $\mathcal{P}_l^s$. Formally we can write them as 
\begin{eqnarray}\label{Pksexis1}
\mathcal{P}_k^s &=& \frac{1}{((k-1)!)^2} \sum_{1,...,k,1',...,k'}
\left|n_k^s(1,..,k)\right> \nonumber \\ && 
\times K_k^s(1,..,k;1',...,k') \left<n_k^s(1',...,k')\right|
\end{eqnarray}
Here $K_k^s$ is the inverse of the $F_k^s(1,...,k;1',...,k')$,
\begin{eqnarray}\label{Fksdef}
F_k^s(1,...,k;1',...,k')
&=& \left<n_k^s(1,...,k)|n_k^s(1',...,k')\right> \nonumber \\
&\equiv & \left<n_k^s(1,...,k)n_k^{s*}(1',...,k')\right>,\;\;\;
\end{eqnarray}
\begin{eqnarray}\label{Pksexis2}
\lefteqn{
\frac{1}{(k-1)!}\sum_{1'',...,k''} \left<n_k^s(1,...,k)n_k^{s*}(1'',...,k'')\right>}
\nonumber \\ && \times
K_k^s(1'',...,k'';1',...,k') = I_k^s(1,...,k;1',...,k').\;\;\;
\end{eqnarray}
In Eq. (\ref{Pksexis2}) $I_k^s(1,...,k;1',...,k')$ is an identity defined as
\begin{equation}\label{Iksdef}
I_k^s(1,...,k;1',...,k') = \sum_{\wp(2',...,k')} \delta_{1,1'}...\delta_{k,k'}
\end{equation}
where $\wp(2',...,k')$ denotes a permutation of the arguments $2',...,k'$, and the sum
in Eq. (\ref{Iksdef}) is over $(k-1)!$ distinct permutations.

The question of the existence of projection operators $\mathcal{P}_k^s$
is equivalent to that of the existence of functions $K_k^s$. Here, we assume
here that these functions exists and we leave the proof of this fact for a future study
(such a proof probably can be done following the analysis presented in
the Appendix B of Ref. \cite{H1}).

It should be noted that the bases $\left|n_k\right>$ and $\left|n_k^s\right>$
are not independent. For example,
\begin{equation}\label{ex1n1n1s}
\left<n_1^s(\mathbf{k}_1) | n_1(\mathbf{k}_2)\right> = S(k_1)
\delta_{\mathbf{k}_1,\mathbf{k}_2}.
\end{equation}
However, it is easy to see that
\begin{equation}\label{ex2n1n1s}
\left<n_k^s | n_l\right> = 0 
\;\;\; \mathrm{for} \;\;\; k < l.
\end{equation}

\section{Exact, formal equations of motion}\label{eofm}

We start with a formal expression for the time-dependent correlation function
of a $k$-particle density at time $t$ and an $l$-particle density at time $0$,
\begin{equation}\label{timecorr0}
\left<N_k \exp\left(\Omega t\right) N_l^*\right>.
\end{equation}
Here $\Omega$ denotes the Smoluchowski operator,
\begin{equation}\label{Smol}
\Omega = D_0 \sum_{i=1}^N \nabla_i\cdot
\left(\nabla_i-\beta\mathbf{F}_i\right),
\end{equation}
where $D_0$ is the diffusion coefficient of an isolated Brownian
particle, $\nabla_i$ denotes a partial derivative 
with respect to $\mathbf{r}_i$,
\begin{equation}\label{nabla}
\nabla_i = \frac{\partial}{\partial \mathbf{r}_i},
\end{equation}
$\beta=1/(k_B T)$ with $k_B$ being the Boltzmann constant, 
and $\mathbf{F}_i$ denotes a force acting on particle $i$,
\begin{equation}\label{force}
\mathbf{F}_i = \sum_{j\neq i} \mathbf{F}_{ij} = - \sum_{j\neq i} \nabla_i V(r_{ij}).
\end{equation} 
with $V(r)$ being the inter-particle potential. Finally, in expression (\ref{timecorr0})
the equilibrium distribution
stands to the right of the quantity being averaged and the Smoluchowski operator, 
and all other operators act on it as well as on everything else 
(unless parentheses indicate otherwise). 

The orthogonalized many-particle densities $n_k$ are linear combinations of densities
$N_k$ and thus we can easily define the following time-dependent correlation functions, 
\begin{equation}\label{timecorr1}
\left<n_k \exp\left(\Omega t\right) n_l^*\right>.
\end{equation}

As emphasized in Ref. \cite{H2}, the advantage of dealing with time-dependent 
correlation functions (\ref{timecorr1}) is that the initial condition is diagonal,
\textit{i.e.} 
\begin{equation}\label{timecorr1t0}
\left<n_k n_l^*\right> = 0 \;\;\; \mathrm{if} \;\;\; k \neq l.
\end{equation}
Another advantage of using 
functions (\ref{timecorr1}) is that in equations of motion
for $\left<n_k \exp\left(\Omega t\right) n_l^*\right>$ bare interactions 
(\textit{i.e.} forces $\mathbf{F}_i$, $i=1,...,N$) 
are automatically renormalized by equilibrium correlation functions.

To derive a hierarchy of equations of motion for correlation functions
(\ref{timecorr1}) we follow Andersen \cite{H2} 
and ascribe the time-dependence to vectors 
$\left| n_k\right>$. Explicitly, 
$\left| n_k(t)\right>$ is defined as the vector associated
with 
\begin{equation}\label{nktdef}
n_k(\mathbf{k}_1,...,\mathbf{k}_k;t) \equiv 
\exp\left(\Omega^{\dagger} t\right) n_k(\mathbf{k}_1,...,\mathbf{k}_k)
\end{equation}
where $\Omega^{\dagger}$ denotes the adjoint Smoluchowski operator,
\begin{equation}\label{Smoladj}
\Omega^{\dagger} = D_0 \sum_{i=1}^N 
\left(\nabla_i+\beta\mathbf{F}_i\right)\cdot\nabla_i.
\end{equation}
It should be emphasized that the adjoint operator $\Omega^{\dagger}$ 
acts only on the densities.

We decompose the time derivative of $\left|n_k(t)\right>$
into a linear combination of $\left|n_l(t)\right>$,
\begin{eqnarray}\label{timeder}
\lefteqn{\frac{\partial}{\partial t} \left|n_k(1,..,k;t)\right> \equiv 
\left|\Omega^{\dagger}n_k(1,...,k;t)\right>} \nonumber \\ 
&& = \sum_{l=0}^{\infty} 
\frac{1}{l!}\sum_{1,...,l} Q_{kl}(1,...,k;1,...,l) \left|n_l(1,...,l;t)\right>.
\end{eqnarray}
The formulas for the coefficients $Q_{kl}$ can be obtained in a number of ways
(see, \textit{e.g.}, Ref. \cite{H2}). The result is
\begin{eqnarray}\label{Q1}
Q_{kl}(1,...,k;1,...,l) &=& \frac{1}{l!}\sum_{1',...,l'}
\left<n_k(1,...,k) \Omega n_l^*(1',...,l')\right> \nonumber \\ 
&& \times K_l(1',...,l';1,...,l).
\end{eqnarray}

Next, we analyze matrix elements of the Smoluchowski operator,
$\left< n_k \Omega n_l^* \right>$.
Since all the particles are the same and the equilibrium
distribution is symmetric with respect to the particle exchange, we can re-write
matrix element $\left< n_k \Omega n_l^* \right>$ in the following way
\begin{eqnarray}\label{matrixel1}
\lefteqn{\left< n_k(\mathbf{k}_1,...,\mathbf{k}_k) \Omega 
n_l^*(\mathbf{q}_1,...,\mathbf{q}_l) 
\right> = } \nonumber \\ 
&& -D_0 N \left< \left(\nabla_1  n_k(\mathbf{k}_1,...,\mathbf{k}_k)\right)\cdot
\left(\nabla_1 n_l^*(\mathbf{q}_1,...,\mathbf{q}_l)\right)\right>,\;\;\;
\end{eqnarray}
where, as emphasized by the parentheses, derivatives 
$\nabla_1$ act only on the densities. 

It is clear that $\nabla_1  n_k(\mathbf{k}_1,...,\mathbf{k}_k)$
is a linear combination of $n_m^s(\mathbf{q}_1,...,\mathbf{q}_m)$ with $m \le k$. 
This allows us to insert projection operators $\mathcal{P}_m^s$ into
expression (\ref{matrixel1}) for matrix element $\left< n_k \Omega n_l^* \right>$,
\begin{eqnarray}\label{matirxel2}
\lefteqn{\left< n_k(\mathbf{k}_1,...,\mathbf{k}_k) \Omega 
n_l^*(\mathbf{q}_1,...,\mathbf{q}_l) 
\right> = -D_0 N} \\ \nonumber  && 
\times \sum_{m=0}^{\mathrm{min}(k,l)} 
\left< \left(\nabla_1  n_k(\mathbf{k}_1,...,\mathbf{k}_k)\right)
\mathcal{P}_m^s \cdot
\left(\nabla_1 n_l^*(\mathbf{q}_1,...,\mathbf{q}_l)\right)\right>.
\end{eqnarray}

Finally, if $k \ge m$ then, unless $k=m$ or $k=m+1$, 
\begin{equation}\label{matrixel2}
\left< \left(\nabla_1 n_k(\mathbf{k}_1,...,\mathbf{k}_k)\right) 
n_m^{s*}(\mathbf{q}_1,...,\mathbf{q}_m)\right> = 0.
\end{equation}
Eq. (\ref{matrixel2}) follows from integrating
$\left< \left(\nabla_1 n_k\right)n_m^{s*}\right>$ by parts and then using the fact 
that $n_k$ is orthogonal to all $n_l^s$ for $k>l$.
As a consequence of Eq. (\ref{matrixel2}), the only non-vanishing matrix
elements of the Smoluchowski operator are $\left<n_k \Omega n_{k+1}^*\right>$,
$\left<n_{k+1} \Omega n_k^*\right>$ and
$\left<n_k \Omega n_k^*\right>$:
\begin{widetext}
\begin{eqnarray}\label{kkp12}
\lefteqn{\left< n_k(\mathbf{k}_1,...,\mathbf{k}_k) \Omega 
n_{k+1}^*(\mathbf{q}_1,...,\mathbf{q}_{k+1}) 
\right> =} \nonumber \\ &&  iD_0 N 
\sum_{i=1}^k
\mathbf{k}_i \cdot \left< n_k^s(\mathbf{k}_i,\mathbf{k}_2,...,\mathbf{k}_{i-1},
\mathbf{k}_{i+1},...,\mathbf{k}_k)
\left(\nabla_1 n_{k+1}^*(\mathbf{q}_1,...,\mathbf{q}_{k+1})\right)\right>,
\end{eqnarray}
\begin{eqnarray}\label{kp1k2}
\lefteqn{\left< n_k(\mathbf{k}_1,...,\mathbf{k}_{k+1}) \Omega 
n_k^*(\mathbf{q}_1,...,\mathbf{q}_k) 
\right> =} \nonumber \\ &&  -iD_0 N  \sum_{i=1}^k
\left< \left(\nabla_1  n_{k+1}(\mathbf{k}_1,...,\mathbf{k}_{k+1})\right)
n_k^{s*}(\mathbf{q}_i,\mathbf{q}_2,...,\mathbf{q}_{i-1},
\mathbf{q}_{i+1},...,\mathbf{q}_k)\right> \cdot  \mathbf{q}_i,
\end{eqnarray}
\begin{eqnarray}\label{kk2}
\lefteqn{\left< n_k(\mathbf{k}_1,...,\mathbf{k}_k) \Omega 
n_k^*(\mathbf{q}_1,...,\mathbf{q}_k) 
\right> =} \nonumber \\ &&  -D_0 N
\sum_{i,j=1}^k \mathbf{k}_i\cdot\mathbf{q}_j
\left< n_k^s(\mathbf{k}_i,\mathbf{k}_2,...,\mathbf{k}_{i-1},
\mathbf{k}_{i+1},...,\mathbf{k}_k) 
n_k^{s*}(\mathbf{q}_j,\mathbf{q}_2,...,\mathbf{q}_{j-1},
\mathbf{q}_{j+1},...,\mathbf{q}_k)\right> \nonumber \\
&& -D_0 N\left< \left(\nabla_1  n_k(\mathbf{k}_1,...,\mathbf{k}_k)\right)
\mathcal{P}_{k-1}^s \left(\nabla_1 n_k^*(\mathbf{q}_1,...,\mathbf{q}_k)\right)\right>.
\end{eqnarray}
\end{widetext}
One should note that the diagonal matrix element $\left<n_k\Omega n_k^*\right>$ 
consists of two different parts. This decomposition, which appears here in a very
natural way, will lead to the emergence of an irreducible memory function.

To derive the formulas for coefficients $Q_{kl}$, 
we contract the expressions for matrix elements
$\left<n_k\Omega n_l^*\right>$ with functions $K_l$. It is obvious that the only
non-vanishing coefficients $Q_{kl}$ are $Q_{k k+1}$, $Q_{k+1 k}$ and $Q_{kk}$.

We are now in a position to write down a hierarchy of equations of motion
for vectors $\left|n_k(1,...,k;t)\right>$, $k\ge 1$. This hierarchy could be 
a starting point for a theory for time-dependent many-particle density 
correlations. In this paper we are only concerned with the time-dependent
single-particle density correlation function, 
$\left<n_1(1;t)n_1^*(1')\right>$. Thus, rather than presenting the most general 
hierarchy, we write
down an equation of motion for $\left<n_1(1;t)n_1^*(1')\right>$,
\begin{eqnarray}\label{exacthierarchy21}
\lefteqn{\frac{\partial}{\partial t} \left<n_1(1;t) n_1^*(1')\right> = 
\sum_{1''} Q_{11}(1;1'') \left<n_1(1'';t)n_1^*(1')\right>} \nonumber \\
&& + \frac{1}{2!} \sum_{1'',2''}Q_{12}(1;1'',2'')\left<n_2(1'',2'';t)n_1^*(1')\right>, 
\end{eqnarray}
and a hierarchy of equations of motion for functions
that couple to $\left<n_1(1;t)n_1^*(1')\right>$, \textit{i.e.} 
time-dependent many-particle correlations 
$\left<n_k(1,...,k;t)n_1^*(1')\right>$, $k>1$,
\begin{widetext}
\begin{eqnarray}\label{exacthierarchy2}
\lefteqn{\frac{\partial}{\partial t} \left<n_k(1,...,k;t)n_1^*(1')\right> =}
\nonumber \\ && 
\frac{1}{(k-1)!}\sum_{1'',...,(k-1)''} Q_{k k-1}(1,...,k;1'',...,(k-1)'') 
\left<n_{k-1}(1'',...,(k-1)'';t)n_1^*(1')\right> \nonumber \\ 
&+& \frac{1}{k!}\sum_{1'',...,k''}Q_{kk}(1,...,k;1'',...,k'')
\left<n_k(1'',...,k'';t)n_1^*(1')\right> \nonumber \\
&+& \frac{1}{(k+1)!} \sum_{1'',...,(k+1)''} Q_{k k+1}(1,...,k;1'',...,(k+1)'')
\left<n_{k+1}(1'',...,(k+1)'';t)n_1^*(1')\right>.\;\;\;
\end{eqnarray}
\end{widetext}

The hierarchy (\ref{exacthierarchy21}--\ref{exacthierarchy2}) is the main
formal result of this paper. One could now follow Andersen \cite{H2} and use
Eqs. (\ref{exacthierarchy21}--\ref{exacthierarchy2}) as a starting point 
for a formally exact diagrammatic approach. Here, we follow a different 
route: first we approximate vertices $Q_{kl}$ and then we formulate
a diagrammatic approach.

Before introducing approximations, let us comment on general structure 
of Eqs. (\ref{exacthierarchy21}--\ref{exacthierarchy2}). First, 
a given correlation function $\left<n_k(t) n_1^*\right>$ couples,
via equations of motion, to $\left<n_{k-1}(t) n_1^*\right>$ (except for 
$k=1$) and $\left<n_{k+1}(t) n_1^*\right>$. Second, the initial condition for
this hierarchy is very simple, 
\begin{equation}\label{timecorr1t01}
\left<n_k(t=0) n_1^*\right> = 0 \;\;\; \mathrm{if} \;\;\; k > 1.
\end{equation}
Thus, in a hierarchy of integral equations that is equivalent to 
Eqs. (\ref{exacthierarchy21}--\ref{exacthierarchy2}), 
and in an iterative solution of this hierarchy, there are no
terms related to $t=0$ correlations except for 
$\left<n_1(t=0) n_1^*\right> \equiv \left<n_1 n_1^*\right>$.
Third, it can easily be shown that the vertices $Q_{kl}$ can be expressed
in terms of equilibrium correlation functions. Thus, bare interactions
present in a hierarchy of equations of motion for correlation functions
$\left<N_k(t) N_l\right>$ have been renormalized \cite{H2}. In particular, within a
simple approximation discussed in the next section, 
the bare force is replaced by a derivative of a direct correlation function.

\section{Approximate equations of motion: lowest order cluster
expansion terms}\label{eofmapprox}

Vertices $Q_{kl}$ that enter into the 
exact, formal equations of motion (\ref{exacthierarchy21}--\ref{exacthierarchy2})
can be expressed in terms of equilibrium correlation functions. In general, exact 
expressions for higher order vertices include higher order correlation
functions, \textit{i.e.} correlation functions beyond the pair correlation
function $g_2(r)$. 
Such higher order correlation functions are not readily available
and are usually approximated and/or neglected once formal expressions 
for time-dependent functions of interest have been derived. Here, we follow an
alternative route: we approximate
vertices $Q_{kl}$ before deriving a diagrammatic expansion. 

A complete cluster expansion of vertices $Q_{kl}$  can be performed following
Sec. II and Appendix A of Ref. \cite{H3}. We only give expressions for
the lowest order terms in the complete cluster expansion. 
To get these terms it is sufficient to
retain only the lowest order terms in the cluster expansions of the matrix elements
(\ref{kkp12}-\ref{kk2}) and of functions $K_l$. The analysis is straightforward
albeit the intermediate formulas are rather long. 
We need the lowest order cluster expansion terms for 
$\left< n_k^s \left(\nabla_1 n_{k+1}^*\right)\right>$ (and its complex conjugate), 
$\left< n_k^s n_k^{s*}\right>$ and $K_k$. Including only the lowest 
order cluster expansion terms, the first quantity is given by the following
expression
\begin{eqnarray}\label{kkpclexp}
\lefteqn{\left< n_k^s(\mathbf{k}_i,\mathbf{k}_2,...,\mathbf{k}_k)
\left(\nabla_1 n_{k+1}^*(\mathbf{q}_1,...,\mathbf{q}_{k+1})\right)\right> =}
\nonumber \\
&& \sum_{l>j=1}^{k+1} 
\sum_{\wp(\mathbf{q}_1,...,\mathbf{q}_{k+1}[\mathbf{q}_j,\mathbf{q}_l])}
\left< n_1^s(\mathbf{k}_i) \left(\nabla_1 
n_2^*(\mathbf{q}_j,\mathbf{q}_l)\right)\right> \nonumber \\
&& \times 
\left<\prod_{m=1\atop m\neq i }^k n_1(\mathbf{k}_m)
\prod_{n=1\atop j\neq n \neq l }^{k+1} n_1^*(\mathbf{q}_n)\right>^{fac}.
\end{eqnarray}
Here the notation $\mathbf{q}_1,...,\mathbf{q}_{k+1}[\mathbf{q}_j,\mathbf{q}_l]$
means remove $\mathbf{q}_j$ and $\mathbf{q}_l$ from the
preceding list and thus 
$\wp(\mathbf{q}_1,...,\mathbf{q}_{k+1}[\mathbf{q}_j,\mathbf{q}_l])$
denotes a permutation of wavevectors $\mathbf{q}_i$, $1\le i \le k+1$, 
$i\neq j$ and $i\neq l$. 
Finally, in Eq. (\ref{kkpclexp}) the following shorthand notation is used,
\begin{eqnarray}
\lefteqn{\left<n_1(\mathbf{k}_1)...n_1(\mathbf{k}_k) 
n_1^*(\mathbf{q}_1)...n_1^*(\mathbf{q}_k) \right>^{fac} \equiv} \nonumber \\ && 
\left<n_1(\mathbf{k}_1) n_1^*(\mathbf{q}_1)\right> ... 
\left<n_1(\mathbf{k}_k) n_1^*(\mathbf{q}_k)\right>.
\end{eqnarray}
The second quantity, $\left< n_k^s n_k^{s*}\right>$, is given by 
\begin{eqnarray}\label{kkclexp}
\lefteqn{\left< n_k^s(\mathbf{k}_i,\mathbf{k}_2,...,\mathbf{k}_{i-1},
\mathbf{k}_{i+1},...,\mathbf{k}_k) \right.} \nonumber \\ &&  \times
\left. n_k^{s*}(\mathbf{q}_j,\mathbf{q}_2,...,\mathbf{q}_{j-1},
\mathbf{q}_{j+1},...,\mathbf{q}_k)\right> = \nonumber \\ &&  
\sum_{\wp(\mathbf{q}_1,...,\mathbf{q}_k[\mathbf{q}_j])}
\left<n_1^s(\mathbf{k}_i) n_1^{s*}(\mathbf{q}_j)\right> \nonumber \\
&& \times
\left<\prod_{m=1\atop m\neq i }^k n_1(\mathbf{k}_m)
\prod_{n=1\atop n \neq j }^kn_1^*(\mathbf{q}_n)\right>^{fac},
\end{eqnarray}
where the notation $\mathbf{q}_1,...,\mathbf{q}_{k+1}[\mathbf{q}_j]$
means remove $\mathbf{q}_j$ from the preceding list.
Finally, including only the lowest 
order cluster expansion terms, $K_l$ has the following simple form
\begin{eqnarray}\label{Kclexp}
\lefteqn{K_k(\mathbf{k}_1,...,\mathbf{k}_k;\mathbf{q}_1,...,\mathbf{q}_k) =}
\nonumber \\ && \sum_{\wp(\mathbf{q}_1,...,\mathbf{q}_k)}
K_1(\mathbf{k}_1;\mathbf{q}_1)...K_1(\mathbf{k}_k;\mathbf{q}_k).
\end{eqnarray}

We substitute expressions (\ref{kkclexp}-\ref{Kclexp}) into the formulas
for vertices $Q_{kk+1}$, $Q_{k+1k}$, and $Q_{kk}$ and, after some calculations, 
we obtain    
\begin{widetext}
\begin{eqnarray}
\label{Qkkp11}
Q_{kk+1}(\mathbf{k}_1,...,\mathbf{k}_k;\mathbf{q}_1,...,\mathbf{q}_{k+1})&=&
iD_0 N\sum_{i=1}^k \sum_{l>j=1}^{k+1}
\sum_{\mathbf{q}'_j,\mathbf{q}'_l}
\mathbf{k}_i\cdot \left< n_1^s(\mathbf{k}_i) \left(\nabla_1 
n_2^*(\mathbf{q}'_j,\mathbf{q}'_l)\right)\right> 
\\ \nonumber && \times 
K_1(\mathbf{q}'_j;\mathbf{q}_j) K_1(\mathbf{q}'_l;\mathbf{q}_l) 
I_{k-1}(\mathbf{k}_1,...,\mathbf{k}_k[\mathbf{k}_i]|\mathbf{q}_1,...,
\mathbf{q}_{k+1}[\mathbf{q}_j,\mathbf{q}_l]),
\end{eqnarray}
\begin{eqnarray}\label{Qkp1k1}
Q_{k+1k}(\mathbf{k}_1,...,\mathbf{k}_{k+1};\mathbf{q}_1,...,\mathbf{q}_k) &=&  
-iD_0 N \sum_{j>i=1}^{k+1} \sum_{l=1}^k 
\sum_{\mathbf{q}'_l} \left< \left(\nabla_1  n_2(\mathbf{k}_i,\mathbf{k}_j)\right) 
n_1^{s*}(\mathbf{q}'_l)\right> \cdot \mathbf{q}'_l 
\nonumber \\ && \times 
K_1(\mathbf{q}'_l;\mathbf{q}_l) 
I_{k-1}(\mathbf{k}_1,...,\mathbf{k}_{k+1}[\mathbf{k}_i,\mathbf{k}_j]|\mathbf{q}_1,...,
\mathbf{q}_k[\mathbf{q}_l]),
\end{eqnarray}
\begin{eqnarray}\label{Qkk1}
\lefteqn{Q_{kk}(\mathbf{k}_1,...,\mathbf{k}_k;\mathbf{q}_1,...,\mathbf{q}_k) =}
\\ \nonumber &&  -D_0 
N\sum_{i,j=1}^k 
\sum_{\mathbf{q}'_j} \mathbf{k}_i\cdot\mathbf{q}'_j
\left< n_1^s(\mathbf{k}_i) n_1^{s*}(\mathbf{q}'_j)\right> 
K_1(\mathbf{q}'_j;\mathbf{q}_j)
I_{k-1}(\mathbf{k}_1,...,\mathbf{k}_k[\mathbf{k}_i]|\mathbf{q}_1,...,
\mathbf{q}_k[\mathbf{q}_j])
\\ && - D_0 N \sum_{j> i=1}^k  \sum_{m> l=1}^k
\sum_{\wp(\mathbf{q}_1,...,\mathbf{q}_k[\mathbf{q}_l,\mathbf{q}_m])} 
\sum_{\mathbf{q}''_l,\mathbf{q}''_m}
\sum_{\mathbf{q}'_1}\left< \left(\nabla_1  n_2(\mathbf{k}_i,\mathbf{k}_j)\right) 
n^{s*}_1(\mathbf{q}'_1)\right>
\nonumber \\
&& \nonumber \times \left<n^s_1(\mathbf{q}'_1) 
\left(\nabla_1 n_2^*(\mathbf{q}''_l,\mathbf{q}''_m)\right)\right>
K_1(\mathbf{q}''_l;\mathbf{q}_l) 
K_1(\mathbf{q}''_m;\mathbf{q}_m)
I_{k-2}(\mathbf{k}_1,...,\mathbf{k}_k[\mathbf{k}_i,\mathbf{k}_j]|\mathbf{q}_1,...,
\mathbf{q}_k[\mathbf{q}_l,\mathbf{q}_m]).
\end{eqnarray}
\end{widetext}
The right-hand-sides of expressions (\ref{Qkkp11}-\ref{Qkk1}) involve
two-particle correlation function (more precisely, its Fourier transform,
\textit{i.e.} the static structure factor) and function 
$\left< n_1^s\left(\nabla_1 n_2^*\right)\right>$. The exact expression for the 
latter function involves a three-particle correlation function. As is 
customary \cite{Goetze,SL}, we use the convolution approximation for the three-particle
contribution to $\left< n_1^s\left(\nabla_1 n_2^*\right)\right>$, and in this
way we obtain
\begin{eqnarray}\label{conv}
&& \left< n_1^s(\mathbf{k}_1) \left(\nabla_1 
n_2^*(\mathbf{q}_1,\mathbf{q}_2)\right)\right> \nonumber \\
&& =  -i n \delta_{\mathbf{k}_1,\mathbf{q}_1+\mathbf{q}_2} 
\left(c(q_1)\mathbf{q}_1+c(q_2)\mathbf{q}_2\right) S(q_1) S(q_2),
\end{eqnarray}
where $c(k)$ is the direct correlation function, $c(k) = (1-1/S(k))/n$.

Substituting expressions (\ref{Qkkp11}-\ref{Qkk1}) together with approximation
(\ref{conv}) into the formal, exact hierarchy 
(\ref{exacthierarchy21}--\ref{exacthierarchy2})
we get an approximate hierarchy in which all the vertices are expressed in 
terms of the static structure factor and the direct correlation function. 
Before we write down this hierarchy, we
take the thermodynamic limit and replace summations over wavevectors by integrals,
\begin{equation}
\frac{1}{V} \sum_{\mathbf{q}} \to \int \frac{d\mathbf{q}}{(2\pi)^3},
\end{equation}
Kronecker $\delta$s by delta functions,
\begin{equation}
V \delta_{\mathbf{k}\mathbf{q}} \to (2\pi)^3 \delta(\mathbf{k}-\mathbf{q}),
\end{equation}
and identities involving Kronecker $\delta$s by ones involving delta functions,
\begin{eqnarray}
\lefteqn{V^k I_k(1,...,k|1',...,k')  \to
\mathcal{I}_k(1,...,k|1',...,k')\equiv} \nonumber \\ && 
\sum_{\wp(1',...,k')}(2\pi)^3 \delta(1-1')...(2\pi)^3\delta(k-k').
\end{eqnarray}
Also, we introduce the following short-hand notation
\begin{eqnarray}
\mathcal{V}_{12}(\mathbf{k}_1;\mathbf{k}_2,\mathbf{k}_3) &=& 
D_0 (2\pi)^3 \delta(\mathbf{k}_1-\mathbf{k}_2-\mathbf{k}_3) 
\nonumber \\ && \times
\mathbf{k}_1\cdot\left(c(k_2)\mathbf{k}_2+c(k_3)\mathbf{k}_3\right)
\end{eqnarray}
\begin{eqnarray}
\mathcal{V}_{21}(\mathbf{k}_1,\mathbf{k}_2;\mathbf{k}_3) &=&
n D_0 (2\pi)^3 \delta(\mathbf{k}_1+\mathbf{k}_2-\mathbf{k}_3)S(k_1)S(k_2)
\nonumber \\ &\times &
\left(c(k_1)\mathbf{k}_1+c(k_2)\mathbf{k}_2\right)\cdot\mathbf{k}_3 S(k_3)^{-1}
\end{eqnarray}
\begin{eqnarray}
\lefteqn{
\mathcal{V}_{22}(\mathbf{k}_1,\mathbf{k}_2;\mathbf{k}_3,\mathbf{k}_4) = } \nonumber \\ &&
n D_0 (2\pi)^3 S(k_1)S(k_2)\delta(\mathbf{k}_1+\mathbf{k}_2-\mathbf{k}_3-\mathbf{k}_4)
\nonumber \\  &\times & \left(c(k_1)\mathbf{k}_1+c(k_2)\mathbf{k}_2\right) \cdot
\left(c(k_3)\mathbf{k}_3+c(k_4)\mathbf{k}_4\right).
\end{eqnarray}

The final result of this section is 
the following equation of motion for the density correlation function
\begin{eqnarray}\label{notsoexacthierarchy2}
&&\frac{\partial}{\partial t} \left<n_1(\mathbf{k}_1;t)n_1^*(\mathbf{q})\right> =
-\frac{D_0 k_1^2}{S(k_1)}  \left<n_1(\mathbf{k}_1;t)n_1^*(\mathbf{q})\right> 
\nonumber \\ && + 
\frac{1}{2!}\int \frac{d\mathbf{q}_1 d\mathbf{q}_2}{(2\pi)^6}
\mathcal{V}_{12}(\mathbf{k}_1;\mathbf{q}_1,\mathbf{q}_2)
\left<n_2(\mathbf{q}_1,\mathbf{q}_2;t)n_1^*(\mathbf{q})
\right>,\;\;\;\;\;\;
\end{eqnarray}
and a hierarchy of equations for functions $\left<n_k(t) n_1^*\right>$, $k>1$,
\begin{widetext}
\begin{eqnarray}\label{notsoexacthierarchy2a}
\lefteqn{
\frac{\partial}{\partial t} \left<n_k(\mathbf{k}_1,...,\mathbf{k}_k;t)n_1^*(\mathbf{q})
\right> = } \nonumber \\ && 
\frac{1}{(k-1)!} \sum_{j>i=1}^k \sum_{l=1}^{k-1} \int 
\frac{d\mathbf{q}_1 ... d\mathbf{q}_{k-1}}{(2\pi)^{3(k-1)}} 
\mathcal{V}_{21}(\mathbf{k}_i,\mathbf{k}_j;\mathbf{q}_l) \nonumber \\ && \times
\mathcal{I}_{k-2}(\mathbf{k}_1,...,\mathbf{k}_k[\mathbf{k}_i,\mathbf{k}_j]|
\mathbf{q}_1,...,\mathbf{q}_{k-1}[\mathbf{q}_l])
\left<n_{k-1}(\mathbf{q}_1,...,\mathbf{q}_{k-1};t)n_1^*(\mathbf{q})\right> 
\nonumber \\ 
&-& \sum_{i=1}^k \frac{D_0 k_i^2}{S(k_i)} 
\left<n_k(\mathbf{k}_1,...,\mathbf{k}_k;t)n_1^*(\mathbf{q})\right>
\nonumber \\ &-&  
\frac{1}{k!}
\sum_{j>i=1}^k \sum_{m>l=1}^k \int \frac{d\mathbf{q}_1 ... d\mathbf{q}_k}{(2\pi)^{3k}}
\mathcal{V}_{22}(\mathbf{k}_i,\mathbf{k}_j;\mathbf{q}_l,\mathbf{q}_m)
\nonumber \\ && \times
\mathcal{I}_{k-2}(\mathbf{k}_1,...,\mathbf{k}_k[\mathbf{k}_i,\mathbf{k}_j]|
\mathbf{q}_1,...,\mathbf{q}_k[\mathbf{q}_l,\mathbf{q}_m])
\left< n_k(\mathbf{q}_1,...,\mathbf{q}_{k};t)
n_1^*(\mathbf{q})\right>
\nonumber \\   &+& 
\frac{1}{(k+1)!}\sum_{i=1}^k \sum_{l>j=1}^{k+1} 
\int \frac{d\mathbf{q} ... d\mathbf{q}_{k+1}}{(2\pi)^{3(k+1)}}
\mathcal{V}_{12}(\mathbf{k}_i;\mathbf{q}_j,\mathbf{q}_l) \nonumber \\ && \times
\mathcal{I}_{k-1}(\mathbf{k}_1,...,\mathbf{k}_k[\mathbf{k}_i]|\mathbf{q}_1,...,
\mathbf{q}_{k+1}[\mathbf{q}_j,\mathbf{q}_l])
\left<n_{k+1}(\mathbf{q}_1,...,\mathbf{q}_{k+1};t)n_1^*(\mathbf{q})
\right>.
\end{eqnarray}
\end{widetext}

\section{Diagrammatic representation}\label{diagram}

To derive a diagrammatic representation for the time-dependent density correlation
function $\left<n_1(t) n_1\right>$ we replace the hierarchy 
(\ref{notsoexacthierarchy2}-\ref{notsoexacthierarchy2a}) by a hierarchy of 
integral equations. Explicitly, for $t>0$, for the density correlation function we get,
\begin{widetext}
\begin{eqnarray}\label{notsoexactinteq1}
\left<n_1(\mathbf{k}_1;t)n_1^*(\mathbf{q})
\right> &=& 
e^{-\frac{D_0 k_1^2 t}{S(k_1)}} \left<n_1(\mathbf{k}_1)n_1^*(\mathbf{q})\right>   
\\ \nonumber && +  \int_0^t dt' e^{-\frac{D_0 k_1^2 (t-t')}{S(k_1)}} \frac{1}{2!}
\int \frac{d\mathbf{q}_1 d\mathbf{q}_2}{(2\pi)^6} 
\mathcal{V}_{12}(\mathbf{k}_1;\mathbf{q}_1,\mathbf{q}_2) 
\left<n_2(\mathbf{q}_1,\mathbf{q}_2;t')n_1^*(\mathbf{q})\right>,
\end{eqnarray}
and for the higher order functions, $\left<n_k(t) n_1^*\right>$, $k>1$,
we obtain the following hierarchy
\begin{eqnarray}\label{notsoexactinteq2}
\lefteqn{\left<n_k(\mathbf{k}_1,...,\mathbf{k}_k;t)n_1^*(\mathbf{q})
\right> =} \nonumber \\ &&
\frac{1}{(k-1)!} \sum_{j>i=1}^k \sum_{l=1}^{k-1} \int_0^t dt' \prod_{m=1}^k 
e^{-\frac{D_0 k_m^2 (t-t')}{S(k_m)}} 
\int \frac{d\mathbf{q}_1...d\mathbf{q}_{k-1}}{(2\pi)^{3(k-1)}} 
\mathcal{V}_{21}(\mathbf{k}_i,\mathbf{k}_j;\mathbf{q}_l)  \nonumber \\ && \times
\mathcal{I}_{k-2}(\mathbf{k}_1,...,\mathbf{k}_k[\mathbf{k}_i,\mathbf{k}_j]|
\mathbf{q}_1,...,\mathbf{q}_{k-1}[\mathbf{q}_l])
\left<n_{k-1}(\mathbf{q}_1,...,\mathbf{q}_{k-1};t')n_1^*(\mathbf{q})\right> 
\nonumber \\ 
&-& \frac{1}{k!}\sum_{j>i=1}^k \sum_{m>l=1}^k \int_0^t dt' \prod_{n=1}^k 
e^{-\frac{D_0 k_n^2 (t-t')}{S(k_n)}}
\int \frac{d\mathbf{q}_1 ... d\mathbf{q}_k}{(2\pi)^{3k}}
\mathcal{V}_{22}(\mathbf{k}_i,\mathbf{k}_j;\mathbf{q}_l,\mathbf{q}_m) 
\nonumber \\ && \times
\mathcal{I}_{k-2}(\mathbf{k}_1,...,\mathbf{k}_k[\mathbf{k}_i,\mathbf{k}_j]|
\mathbf{q}_1,...,\mathbf{q}_k[\mathbf{q}_l,\mathbf{q}_m])
\left< n_k(\mathbf{q}_1,...,\mathbf{q}_{k};t')
n_1^*(\mathbf{q})\right>
\nonumber \\ &+& 
\frac{1}{(k+1)!}\sum_{i=1}^k \sum_{l>j=1}^k \int_0^t dt' \prod_{m=1}^k 
e^{-\frac{D_0 k_m^2 (t-t')}{S(k_m)}}
\int \frac{d\mathbf{q}_1 ... d\mathbf{q}_{k+1}}{(2\pi)^{3(k+1)}}
\mathcal{V}_{12}(\mathbf{k}_i;\mathbf{q}_j,\mathbf{k}_l) \nonumber \\ && \times
\mathcal{I}_{k-1}(\mathbf{k}_1,...,\mathbf{k}_k[\mathbf{k}_i]|\mathbf{q}_1,...,
\mathbf{q}_{k+1}[\mathbf{q}_j,\mathbf{q}_l]) 
\left<n_{k+1}(\mathbf{q}_1,...,\mathbf{q}_{k+1};t')n_1^*(\mathbf{q})
\right>.
\end{eqnarray}
\end{widetext}
The hierarchy of integral equations 
(\ref{notsoexactinteq1}-\ref{notsoexactinteq2}) can be solved
with respect to (w.r.t.) the time-dependent density correlation function 
$\left<n_1(t) n_1\right>$  by iteration. We can express the
latter function in terms of so-called response function $G(k;t)$ that 
is defined through the following equation
\begin{equation}\label{Gdef}
\theta(t) \left<n_1(\mathbf{k};t)n_1^*(\mathbf{q}) \right> = 
n G(k;t) S(k) (2\pi)^3 \delta(\mathbf{k}-\mathbf{q}).
\end{equation} 
Note that the correlation function $\left<n_1(\mathbf{k};t)n_1^*(\mathbf{q}) \right>$
is diagonal in the wavevector space due to the translational invariance. 
To simplify notation we also introduce bare response function $G_0(k;t)$,
\begin{equation}\label{G0def}
G_0(k;t) = \theta(t) \exp(-D_0 k^2 t/S(k)).
\end{equation}

Iterating (\ref{notsoexactinteq1}-\ref{notsoexactinteq2})
a few times we can easily generate the first few terms of the complete
infinite series
\begin{widetext}
\begin{eqnarray}\label{Giter1}
\lefteqn{G(k;t) = G_0(k;t)} 
\nonumber \\ &&
+  \int dt' dt'' 
\int \frac{d\mathbf{k}_1 d\mathbf{k}_2 d\mathbf{k}_3}{2! 1! (2\pi)^9} G_0(k;t-t') 
\mathcal{V}_{12}(\mathbf{k};\mathbf{k}_1,\mathbf{k}_2) G_0(k_1;t'-t'')G_0(k_2;t'-t'')
\mathcal{V}_{21}(\mathbf{k}_1,\mathbf{k}_2;\mathbf{k}_3)G_0(k_3;t'')
\nonumber \\ &&
-  \int dt' dt'' dt'''
\int \frac{d\mathbf{k}_1 d\mathbf{k}_2 d\mathbf{k}_3 d\mathbf{k}_4 d\mathbf{k}_5}
{2! 2! 1! (2\pi)^{15}}G_0(k;t-t') \mathcal{V}_{12}(\mathbf{k};\mathbf{k}_1,\mathbf{k}_2)
G_0(k_1;t'-t'')G_0(k_2;t'-t'')
\nonumber \\ && \times
\mathcal{V}_{22}(\mathbf{k}_1,\mathbf{k}_2;\mathbf{k}_3,\mathbf{k}_4)
G_0(k_3;t''-t''')G_0(k_4;t''-t''')
\mathcal{V}_{21}(\mathbf{k}_3,\mathbf{k}_3;\mathbf{k}_5) G_0(k_5;t''').
\end{eqnarray}
\end{widetext}
Note that in Eq. (\ref{Giter1}) we do not need restrictions on integrations over time
due to the presence of $\theta$ function in the definition of the
bare response function.

\begin{figure}
\includegraphics[scale=.2]{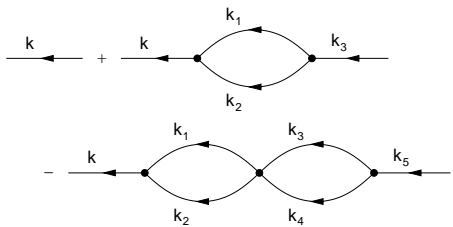}
\caption{Diagrammatic representation of the terms in the series
expansion of the response function $G(k;t)$ showed in Eq. (\ref{Giter1}).} 
\label{f:Giter}
\end{figure}

Terms on the right-hand-side of the above expression, and all other terms in 
the iterative solution of the hierarchy (\ref{notsoexactinteq1}-\ref{notsoexactinteq2}), 
can be represented by diagrams. The diagrammatic rules are as follows: 
\begin{itemize}
\item response function $G(k;t)$:
\includegraphics[scale=.2]{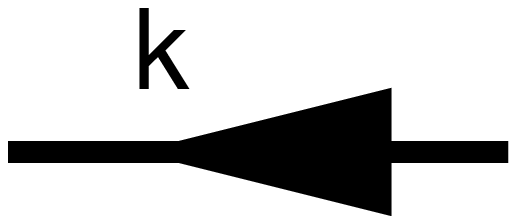}
\item bare response function $G_0(k;t)$:
\includegraphics[scale=.2]{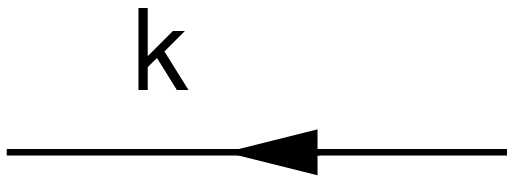}
\item ``left'' vertex $\mathcal{V}_{12}$:
\includegraphics[scale=.2]{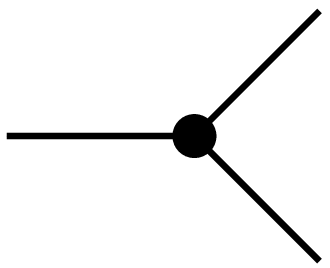}
\item ``right'' vertex $\mathcal{V}_{21}$: 
\includegraphics[scale=.2]{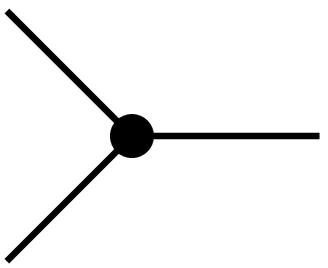}
\item four-leg vertex $\mathcal{V}_{22}$:
\includegraphics[scale=.2]{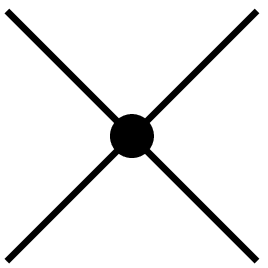}
\item $(2\pi)^3 \delta$ vertex:
\includegraphics[scale=.2]{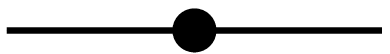}
\end{itemize}
We refer to the leftmost bare response function as 
left root, and to the other bare response functions as bonds.
To calculate a diagram one integrates over all wavevectors (with a $(2\pi)^{-3}$ 
factor for each integration) except the wavevector
corresponding to the left root. Furthermore, one integrates over all intermediate times, 
and divides the result by a product of factorials that follow from 
factorials appearing in hierarchy (\ref{notsoexactinteq1}-\ref{notsoexactinteq2}).
Diagrams with odd and even numbers of $\mathcal{V}_{22}$ vertices contribute with
overall negative and positive sign, respectively. For illustration,  
diagrammatic representation of the right-hand-side of Eq. (\ref{Giter1}) 
is shown in Fig. \ref{f:Giter}.

\begin{figure}
\includegraphics[scale=.2]{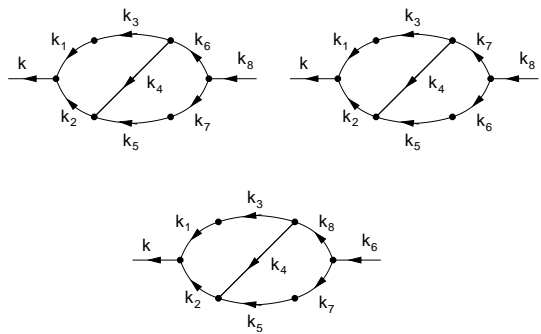}
\caption{According to hierarchy (\ref{notsoexactinteq1}-\ref{notsoexactinteq2})
for a given sequence of wavevectors, only those topologically different 
diagrams that differ by 
permutations of wavevector labels pertaining to the same ``time slice'' are
allowed. Thus, the top diagrams should be included whereas the bottom one should not.} 
\label{f:counting}
\end{figure}

It is very important to note that 
labeled diagrams that occur in the series expansion generated by the iterative 
solution of hierarchy (\ref{notsoexactinteq1}-\ref{notsoexactinteq2}) differ by
a permutation of labels pertaining to the same ``time slice''. For example, 
out of three diagrams showed in Fig. \ref{f:counting} 
only the top two ones enter in the series expansion and
including also the third one would lead to over-counting. Thus, in the following
by topologically different labeled diagrams we mean only those topologically different
diagrams that differ by
a permutation of labels pertaining to the same ``time slice''.

Summarizing, 
we obtain the following diagrammatic representation of the response function:
\begin{eqnarray}\label{Gseries0}
&& G(k;t) = \\ &&  \nonumber 
\text{sum of all topologically different labeled diagrams}  \\ && \nonumber 
\text{with a left root labeled $k$, a right root, 
$G_0$ bonds, } \\ && \nonumber  
\text{$\mathcal{V}_{12}$, $\mathcal{V}_{21}$ and $\mathcal{V}_{22}$
vertices, in which diagrams with}  \\ &&  \nonumber
\text{odd and even numbers of 
$\mathcal{V}_{22}$ vertices contribute}  \\ && \nonumber 
\text{with overall negative and positive sign, respectively.}
\end{eqnarray}

\begin{figure}
\includegraphics[scale=.2]{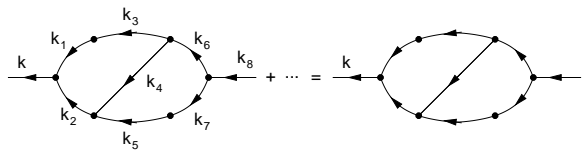}
\caption{24 labeled diagrams resulting from permutations of wavevector labels 
pertaining to their respective ``time slices''
lead to one unlabeled diagram showed on the right. The symmetry
number of this diagram is $S=1$.}
\label{f:labelunlabel}
\end{figure}

Next, we introduce unlabeled diagrams. Bonds in these diagrams, except for the 
left root, are not labeled. Two unlabeled diagrams are topologically
equivalent if there is a way to assign labels to unlabeled bonds so that the
resulting labeled diagrams are topologically equivalent. To evaluate an unlabeled
diagram one assigns labels to unlabeled bonds, evaluates the resulting 
labeled diagram, and then divides the result by a symmetry number of the
diagram (\textit{i.e.} the number of topologically identical labeled diagrams
that can be obtained from a given unlabeled diagram by permutation of
the bond labels). 
It should be appreciated that each unlabeled diagram represents
a number of original, labeled diagrams. For example, the labeled diagram showed 
in Fig. \ref{f:labelunlabel} 
and another 23 similar labeled diagrams (\textit{i.e.} 24 diagrams altogether)
are represented by one unlabeled diagram. 
It can be showed that the diagrammatic series (\ref{Gseries0}) can be replaced 
by a series of topologically different unlabeled diagrams. To prove this fact
one has to follow the proof of  
an analogous transformation from a series of labeled Mayer diagrams to a series of
unlabeled ones \cite{Mayer}.

\begin{figure}
\includegraphics[scale=.2]{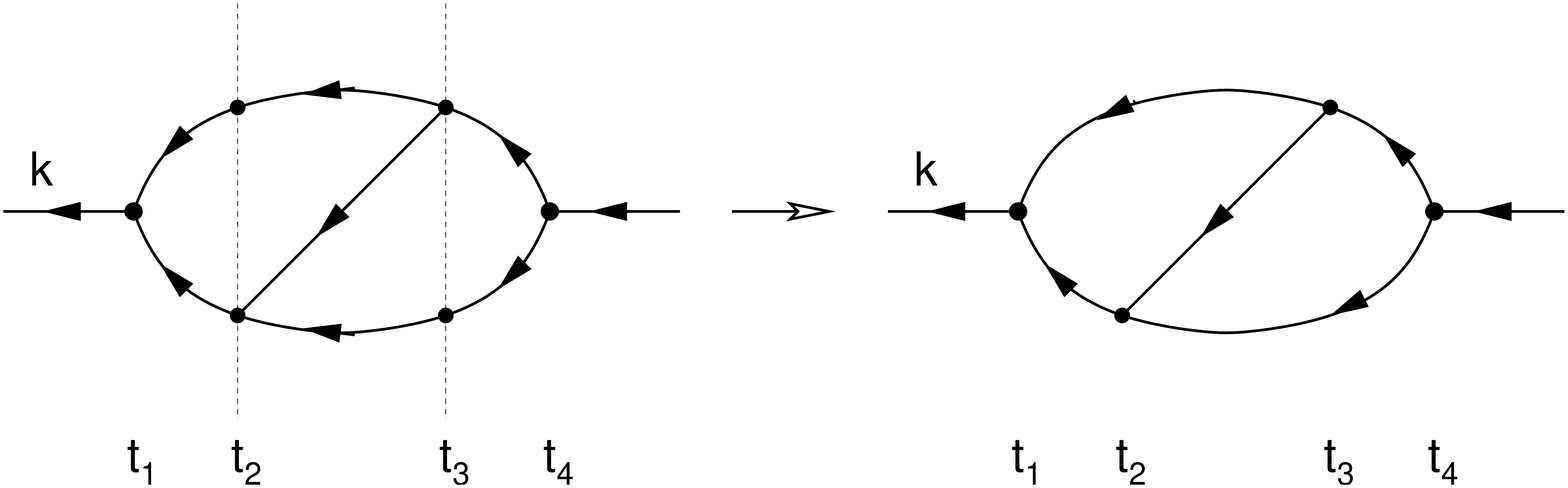}
\caption{Unlabeled diagram on the left is converted into an unlabeled diagram
without two-leg vertices corresponding to $(2\pi)^3\delta$ functions which is 
showed on the right.}
\label{f:nodeltas}
\end{figure}

The unlabeled diagrams can be simplified. 
This can be illustrated on the example of the diagram showed above. 
The value of this diagram is given by the following expression:
\begin{eqnarray}\label{diag21}
&& \int dt_1 dt_2 dt_3 dt_4 \int 
\frac{d\mathbf{k}_1 d\mathbf{k}_2 d\mathbf{k}_3 d\mathbf{k}_4 d\mathbf{k}_5
d\mathbf{k}_6 d\mathbf{k}_7 d\mathbf{k}_8}{(2\pi)^{24}} 
\nonumber \\ && \times G_0(k;t-t_1) 
\mathcal{V}_{12}(\mathbf{k};\mathbf{k}_1,\mathbf{k}_2)
G_0(k_1;t_1-t_2) \nonumber \\ && \times G_0(k_2;t_1-t_2)
\mathcal{V}_{12}(\mathbf{k}_2;\mathbf{k}_4,\mathbf{k}_5)
(2\pi)^3\delta(\mathbf{k}_1-\mathbf{k}_3) \nonumber \\ && \times
G_0(k_3;t_2-t_3) G_0(k_4;t_2-t_3)G_0(k_5;t_2-t_3) \nonumber \\ && \times 
\mathcal{V}_{21}(\mathbf{k}_3,\mathbf{k}_4;\mathbf{k}_6)
(2\pi)^3\delta(\mathbf{k}_5-\mathbf{k}_7) 
G_0(k_6;t_3-t_4) \nonumber \\ && \times 
G_0(k_7;t_3-t_4) \mathcal{V}_{21}(\mathbf{k}_6,\mathbf{k}_7;\mathbf{k}_8)
G_0(k_8;t_9).
\end{eqnarray}
To simplify this formula we first integrate
over $\delta$ functions. Then, 
since $G_0(k_1;t_1-t_2)G_0(k_1;t_2-t_3) = 
G_0(k_1;t_1-t_3)\theta(t_1-t_2)\theta(t_2-t_3)$ and $G_0(k_5;t_2-t_3)G_0(k_5;t_3-t_4) = 
G_0(k_5;t_2-t_4)\theta(t_2-t_3)\theta(t_3-t_4)$,
we can rewrite (\ref{diag21}) in 
the following form
\begin{eqnarray}\label{diag22}
&& \int dt_1 dt_2 dt_3 dt_4 \int 
\frac{d\mathbf{k}_1 d\mathbf{k}_2 d\mathbf{k}_4 d\mathbf{k}_5 d\mathbf{k}_6
d\mathbf{k}_8}
{(2\pi)^{18}} G_0(k;t-t_1) \nonumber \\ && \times 
\mathcal{V}_{12}(\mathbf{k};\mathbf{k}_1,\mathbf{k}_2)
G_0(k_1;t_1-t_3)G_0(k_2;t_1-t_2) \nonumber \\ && \times 
\mathcal{V}_{12}(\mathbf{k}_2;\mathbf{k}_4,\mathbf{k}_5)
G_0(k_4;t_2-t_3) \nonumber \\ && \times G_0(k_5;t_2-t_4)
\mathcal{V}_{21}(\mathbf{k}_1,\mathbf{k}_4;\mathbf{k}_6)
\nonumber \\ && \times 
G_0(k_6;t_3-t_4) \mathcal{V}_{21}(\mathbf{k}_5,\mathbf{k}_7;\mathbf{k}_9)
G_0(k_8;t_4).
\end{eqnarray}
One should note that in the above expression additional $\theta$ functions 
originating from simplifying products of bare response functions have been incorporated
into the remaining bare response functions. 
Diagrammatic interpretation of the above described transformation
is showed in Fig. \ref{f:nodeltas}.

\begin{figure}
\includegraphics[scale=.2]{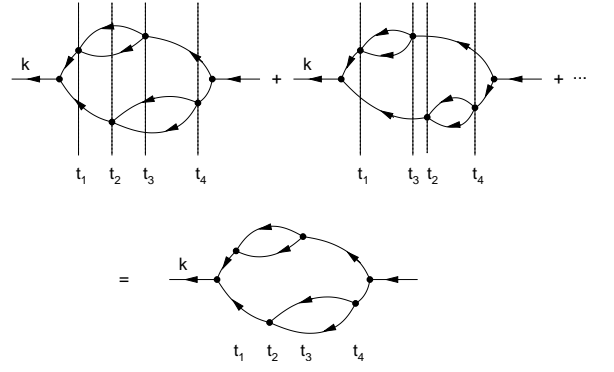}
\caption{Time ordered diagrams corresponding to different time orderings of 
vertices can be re-summed to give one time-unordered diagram.}
\label{f:tunordered}
\end{figure}

For more complicated diagrams after integrating over 
$\delta$ functions we are still left with explicit, specific time ordering of
vertices. For example, for diagrams showed in Fig. \ref{f:tunordered}
we have $t_1\ge t_2\ge t_3\ge t_4$ for the first diagram, 
$t_1\ge t_3\ge t_2\ge t_4$ for the second, etc. We can, however, sum all
such diagrams and obtain a time-unordered diagram for which 
there is no restriction on the ordering of times associated
with different vertices (note that there is always an implicit restriction
due to vanishing of response functions for $t<0$ and thus in the above
example $t_1 \ge t_3$ and $t_2\ge t_4$). 
As a result we obtain a series involving so-called time-unordered diagrams. 
One should note that both integrating over $\delta$ functions and 
replacing sums of time-ordered diagrams by time-unordered ones do not change
the symmetry numbers of diagrams.
In the following we implicitly assume that these transformations
have been performed on all diagrams. Thus, from now on, we will consider 
time-unordered diagrams without $(2\pi)^3\delta$ vertices.

The final result of this section is the following diagrammatic expression for the
response function: 
\begin{eqnarray}\label{Gseries}
&& G(k;t) = \\ &&  \nonumber 
\text{sum of all topologically different diagrams with}  \\ && \nonumber 
\text{a left root labeled $k$, a right root, 
$G_0$ bonds, $\mathcal{V}_{12}$,} \\ && \nonumber  
\text{$\mathcal{V}_{21}$ and $\mathcal{V}_{22}$
vertices, in which diagrams with odd}  \\ &&  \nonumber
\text{and even numbers of 
$\mathcal{V}_{22}$ vertices contribute with}  \\ && \nonumber 
\text{overall negative and positive sign, respectively.}
\end{eqnarray}

\section{Memory functions: reducible and irreducible}\label{mf} 

\begin{figure}
\includegraphics[scale=.2]{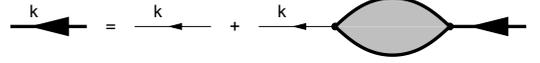}
\caption{Diagrammatic representation of the Dyson equation, Eq. (\ref{Dyson}).}
\label{f:dyson}
\end{figure}

We start with the Dyson equation
\begin{eqnarray}\label{Dyson}
G(k;t) &=& G_0(k;t) + \int dt_1 dt_2 \int \frac{d\mathbf{k}_1}{(2\pi)^3} 
G_0(k;t-t_1) \nonumber \\ && \times \Sigma(\mathbf{k},\mathbf{k}_1;t_1-t_2) G(k_1;t_2),
\end{eqnarray}
where $\Sigma$ is the self energy.
Diagrammatic representation of the Dyson equation is showed in Fig. \ref{f:dyson}.
Due to the translational invariance the self-energy is diagonal in wavevector,
\begin{equation}\label{sediag}
\Sigma(\mathbf{k},\mathbf{k}_1;t) \propto (2\pi)^3 \delta(\mathbf{k}-\mathbf{k}_1).
\end{equation}

It follows from general analysis of the Dyson equation
that the self-energy $\Sigma$ is a sum of diagrams that do not separate 
into disconnected components upon removal of a single bond. 

The memory function can be obtain from $\Sigma$ in the following way.
We note that the diagrams contributing to the self-energy
start with $\mathcal{V}_{21}$ vertex on the right and end with 
$\mathcal{V}_{12}$ vertex on the left. Customarily, to define the memory function
for a Brownian system one factors out parts of these vertices. First, we define
memory matrix $\mathbf{M}$ by factoring out $\mathbf{k}$ from the left  vertex
and $(D_0/S(k_1)) \mathbf{k}_1$ from the right vertex,
\begin{equation}\label{matm}
\Sigma(\mathbf{k},\mathbf{k}_1;t) = 
D_0
\mathbf{k}\cdot\mathbf{M}(\mathbf{k},\mathbf{k}_1;t)\cdot\mathbf{k}_1 S(k_1)^{-1}.
\end{equation}
Due to the translational and rotational invariance $\mathbf{M}$ is
diagonal in the wavevector and longitudinal. Thus we can define memory
function $M$ through the following relation
\begin{equation}\label{matmdiag}
\mathbf{M}(\mathbf{k},\mathbf{k}_1;t) =M(k;t) \hat{\mathbf{k}} \hat{\mathbf{k}}
(2\pi)^3 \delta(\mathbf{k}-\mathbf{k}_1).
\end{equation}
Using Eq. (\ref{matm}) and (\ref{matmdiag}) 
we can obtain the following equation from the Laplace 
transform of the Dyson equation,
\begin{equation}\label{Dysonm}
G(k;z) = G_0(k;z) + G_0(k;z) \frac{D_0 k^2}{S(k)} M(k;z) G(k;z).
\end{equation}
Eq. (\ref{Dysonm}) can be solved w.r.t. response function $G(k;z)$.
Using the definition of bare response function $G_0$ we obtain 
\begin{equation}\label{Gm}
G(k;z) = \frac{1}{z+ \frac{D_0 k^2}{S(k)} \left(1-M(k;z)\right)}.
\end{equation}
Multiplying both sides of the above equation by the static structure
factor and noting that $S(k) G(k;z)=F(k;z)$, where $F$ is the collective
intermediate scattering function, we get the well-known
memory function representation\cite{HK} of $F(k;z)$,
\begin{equation}\label{Fm}
F(k;z) = \frac{S(k)}{z+ \frac{D_0 k^2}{S(k)} \left(1-M(k;z)\right)}.
\end{equation}

To facilitate further discussion it is convenient to 
introduce cut-out vertices corresponding to the following functions:
\begin{eqnarray} \label{V12c}
\mathbf{V}_{12}^c(\mathbf{k}_1;\mathbf{k}_2,\mathbf{k}_3) &=& 
D_0 (2\pi)^3 \delta(\mathbf{k}_1-\mathbf{k}_2-\mathbf{k}_3) 
\nonumber \\ && \times
\left(c(k_2)\mathbf{k}_2+c(k_3)\mathbf{k}_3\right)
\end{eqnarray}
\begin{eqnarray}\label{V21c}
\mathbf{V}_{21}^c(\mathbf{k}_1,\mathbf{k}_2;\mathbf{k}_3) &=&
n (2\pi)^3 \delta(\mathbf{k}_1+\mathbf{k}_2-\mathbf{k}_3)S(k_1)S(k_2)
\nonumber \\ && \times 
\left(c(k_1)\mathbf{k}_1+c(k_2)\mathbf{k}_2\right).
\end{eqnarray}
These vertices are obtained by factoring out $\mathbf{k}_1$ from 
vertex $\mathcal{V}_{12}$ and $(D_0/S(k_3)) \mathbf{k}_3$
from vertex $\mathcal{V}_{21}$.

It should be noted that 
\begin{eqnarray}\label{fact}
&& \mathcal{V}_{22}(\mathbf{k}_1,\mathbf{k}_2;\mathbf{k}_3,\mathbf{k}_4)  
\nonumber \\ && =
\int \frac{d \mathbf{k}'}{(2\pi)^3}
\mathbf{V}_{21}^c(\mathbf{k}_1,\mathbf{k}_2;\mathbf{k}')\cdot
\mathbf{V}_{12}^c(\mathbf{k}';\mathbf{k}_3,\mathbf{k}_4).
\end{eqnarray}
The diagrammatic 
rules for functions $\mathbf{V}_{12}^c$ and $\mathbf{V}_{21}^c$ are as follows:
\begin{itemize}
\item ``left'' cut-out vertex $\mathbf{V}_{12}^c$:
\includegraphics[scale=.2]{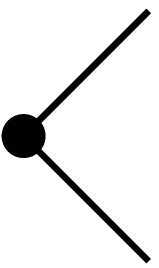}
\item ``right'' cut-out vertex $\mathbf{V}_{21}^c$: 
\includegraphics[scale=.2]{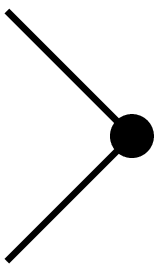}
\end{itemize}
and we refer to wavevector $\mathbf{k}_1$ in 
$\mathbf{V}_{12}^c(\mathbf{k}_1;\mathbf{k}_2,\mathbf{k}_3)$
and $\mathbf{k}_3$ in $\mathbf{V}_{21}^c(\mathbf{k}_1,\mathbf{k}_2;\mathbf{k}_3)$
as roots of these vertices.

\begin{figure}
\includegraphics[scale=.2]{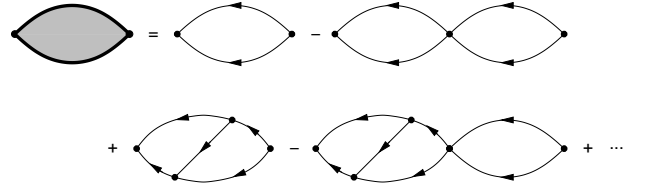}
\caption{The first few diagrams in series expansion for memory matrix $\mathbf{M}$.}
\label{f:mred}
\end{figure}

It follows from the definition of the memory matrix $\mathbf{M}$
that 
\begin{eqnarray}\label{Mredseries}
&& \mathbf{M}(\mathbf{k},\mathbf{k}_1;t) = \\ &&  \nonumber 
\text{sum of all topologically different diagrams which}  \\ && \nonumber 
\text{do not separate into disconnected components}  \\ && \nonumber 
\text{upon removal of a single bond, with vertex $\mathbf{V}_{12}^c$}  \\ && \nonumber 
\text{with root $\mathbf{k}$ on the left and vertex $\mathbf{V}_{21}^c$
with root $\mathbf{k}_1$} \\ && \nonumber  \text{on the right,
$G_0$ bonds, $\mathcal{V}_{12}$, $\mathcal{V}_{21}$ and $\mathcal{V}_{22}$
vertices,}  \\ &&  \nonumber \text{in which diagrams without and even numbers of 
$\mathcal{V}_{22}$}  \\ && \nonumber 
\text{vertices contribute with overall negative and}  \\ && \nonumber 
\text{positive sign, respectively.}
\end{eqnarray}
The first few diagrams in the series for $\mathbf{M}$ are showed in Fig. \ref{f:mred}.

Now, one can understand the need for an irreducible memory function \cite{CHess}. 
The series expansion for $\mathbf{M}$ consists of diagrams that are
one-propagator irreducible (\textit{i.e.} diagrams that do not separate into 
disconnected components upon removal of a single bond) but not all of these diagrams are 
completely one-particle irreducible. Some of the diagrams contributing to $\mathbf{M}$
separate into disconnected
components upon removal of $\mathcal{V}_{22}$ vertex (and bonds attached to this
vertex). The examples of such diagrams are the second and
the fourth diagrams on the right-hand-side of the equality sign in Fig. \ref{f:mred}.

\begin{figure}
\includegraphics[scale=.2]{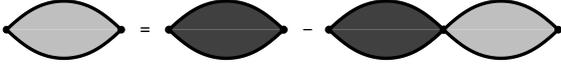}
\caption{Memory matrix $\mathbf{M}$ can be represented as a sum of
$\mathbf{M}^{irr}$ and all other diagrams. The latter diagrams can be re-summed and
it is easy to see that as a result we get the second diagram at the right-hand-side.}
\label{f:mredirr}
\end{figure}

We define the irreducible memory matrix $\mathbf{M}^{irr}$ 
as a sum of only those diagrams
in the series for $\mathbf{M}$ that do not separate into disconnected components
upon removal of a single $\mathcal{V}_{22}$ vertex. Diagrammatically, 
we can represent memory matrix $\mathbf{M}$ as a sum of
$\mathbf{M}^{irr}$ and all other diagrams. The latter diagrams can
be re-summed as showed in Fig. \ref{f:mredirr}. 
Using Eq. (\ref{fact}), we can introduce an additional integration over a
wavevector and then we see that the diagrammatic equation showed in Fig. \ref{f:mredirr}
corresponds to the following equation,
\begin{eqnarray}\label{mredirr} 
&& \mathbf{M}(\mathbf{k},\mathbf{k}_1;t)
= \mathbf{M}^{irr}(\mathbf{k},\mathbf{k}_1;t)
\\ \nonumber && - \int dt_1 \int \frac{d\mathbf{k}_2}{(2\pi)^3}
\mathbf{M}^{irr}(\mathbf{k},\mathbf{k}_2;t-t_1)\cdot
\mathbf{M}(\mathbf{k}_2,\mathbf{k}_1;t_2)
\end{eqnarray}
Again, we use translational and rotational invariance to introduce the 
irreducible memory function $M^{irr}$,
\begin{equation}\label{mirr}
\mathbf{M}^{irr}(\mathbf{k},\mathbf{k}_1;t) = M^{irr}(k;t) \hat{\mathbf{k}} 
\hat{\mathbf{k}} (2\pi)^3 \delta(\mathbf{k}-\mathbf{k}_1),
\end{equation}
Taking Laplace transform of Eq. (\ref{mredirr}) and then 
using Eq. (\ref{mirr}) we obtain
\begin{equation}\label{mredirr2}
M(k;z) = M^{irr}(k;z) - M^{irr}(k;z) M(k;z).
\end{equation}
This equation can be solved w.r.t. memory function $M$.  Substituting the solution
into Eq. (\ref{Fm}) we obtain a representation of the intermediate
scattering function in terms of the irreducible memory function,
\begin{equation}\label{Fmirr}
F(k;z) = S(k) G(k;z) = \frac{S(k)}{z+ \frac{D_0 k^2}{S(k)\left(1+M^{irr}(k;z)\right)}}.
\end{equation}
Eq. (\ref{Fmirr}) was first derived by Cichocki and Hess\cite{CHess}. 
It has been used as a starting point for the development of mode-coupling
approximations for both equilibrium \cite{SL} and driven Brownian systems \cite{FC}. 

Diagrammatically, 
\begin{eqnarray}\label{Mirrseries}
&& \mathbf{M}(\mathbf{k},\mathbf{k}_1;t) = \\ &&  \nonumber 
\text{sum of all topologically different diagrams which}  \\ && \nonumber 
\text{do not separate into disconnected components}  \\ && \nonumber 
\text{upon removal of a single bond or a single $\mathcal{V}_{22}$}  
\\ && \nonumber 
\text{vertex, with vertex $\mathbf{V}_{12}^c$ with root $\mathbf{k}$ on the left}  
\\ && \nonumber 
\text{and vertex $\mathbf{V}_{21}^c$
with root $\mathbf{k}_1$ on the right,
$G_0$ bonds, } \\ && \nonumber  
\text{$\mathcal{V}_{12}$, $\mathcal{V}_{21}$ and $\mathcal{V}_{22}$
vertices, in which diagrams with}  \\ &&  \nonumber
\text{odd and even numbers of 
$\mathcal{V}_{22}$ vertices contribute}  \\ && \nonumber 
\text{with overall negative and positive sign, respectively.}
\end{eqnarray}

\begin{figure}
\includegraphics[scale=.2]{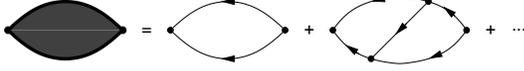}
\caption{The first few diagrams in series expansion for irreducible 
memory matrix $\mathbf{M}^{irr}$.}
\label{f:mirr}
\end{figure}

The first few diagrams in the series for $\mathbf{M}^{irr}$ are showed in 
Fig. \ref{f:mirr}.

\section{Mode-coupling approximation}\label{mct}

\begin{figure}
\includegraphics[scale=.2]{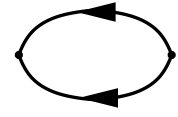}
\caption{Re-summation of diagrams that separate into two disconnected 
components upon removal of the $\mathbf{V}_{12}^c$ and $\mathbf{V}_{21}^c$ vertices
leads to a one-loop diagram with $G$ bonds.}
\label{f:mct}
\end{figure}

The simplest re-summation of the series (\ref{Mirrseries}) includes diagrams
that separate into two disconnected components upon removal of the 
$\mathbf{V}_{12}^c$ and $\mathbf{V}_{21}^c$ vertices. It is easy to see that 
in such diagrams each of these components is a part of the series for the
response function $G$. Summing all such diagrams we get a one-loop diagram
(\textit{i.e.} the first diagram showed on the right-hand-side in Fig. \ref{f:mirr}) 
but with $G_0$ bonds replaced by $G$ bonds, see Fig. \ref{f:mct}.

As a result of this re-summation we get one-loop self-consistent approximation
for the memory matrix, 
\begin{eqnarray}\label{mmct}
\mathbf{M}^{irr}_{one-loop}(\mathbf{k},\mathbf{k}_1;t) &=& 
\int \frac{d\mathbf{k}_2 d\mathbf{k}_3}{2(2\pi)^6} 
\mathbf{V}_{12}^c(\mathbf{k};\mathbf{k}_2,\mathbf{k}_3) \\ \nonumber && \times
G(k_2;t) G(k_3;t) \mathbf{V}_{21}^c(\mathbf{k}_2,\mathbf{k}_3;\mathbf{k}_1).
\end{eqnarray}
The factor 2 in the denominator is the symmetry number of the single-loop diagram. 

Using explicit expressions (\ref{V12c}-\ref{V21c}) for the cut-out vertices
we easily show that (\ref{mmct}) leads to the following expression for the 
irreducible memory function
\begin{eqnarray}\label{mmctexplicit}
\lefteqn{M^{irr}_{one-loop}(k;t) = } 
\\ && 
\frac{n D_0}{2} \int \frac{d\mathbf{k}_1}{(2\pi)^3} 
\left(c(k_1)\hat{\mathbf{k}}\cdot\mathbf{k}_1+
c(|\mathbf{k}-\mathbf{k}_1|)\hat{\mathbf{k}}\cdot(\mathbf{k}-\mathbf{k}_1)\right)^2 
\nonumber \\ && \times
S(k_1)S(|\mathbf{k}-\mathbf{k}_1|)G(k_1;t) G(|\mathbf{k}-\mathbf{k}_1|;t)
\equiv M^{irr}_{MCT}(k;t)
\nonumber
\end{eqnarray}
As indicated above, the one-loop self-consistent approximation coincides with
the mode-coupling approximation, \textit{i.e.} both approximations result
in exactly the same expression for the irreducible memory function.

Expression (\ref{mmctexplicit}) was first derived using a projection operator
approach \cite{SL}. Subsequently, it was also derived using a field theory 
version \cite{KM} of a dynamic density functional theory of Kawasaki \cite{Kawasaki}. 
Later, it was noticed \cite{field1} that
the latter derivation was incompatible with the fluctuation-dissipation
theorem. Recently, there appeared two new field-theoretical derivations 
of the mode-coupling theory for Brownian systems \cite{field2,field3}. 
Only one of these derivations \cite{field3} 
leads to expression (\ref{mmctexplicit}) that was 
originally derived using projection operator method. The other derivation \cite{field2} 
results in a equation that has the same structure as (\ref{mmctexplicit}) but involves 
different vertices.

\section{Discussion}\label{discussion}

We have presented a diagrammatic formulation of a theory for the time dependence of 
density fluctuations in equilibrium systems of interacting Brownian particles.
We have analyzed the series expansion for the time-dependent 
response function and have obtained diagrammatic expressions for both the
memory function and the irreducible memory function. The one-loop self-consistent 
approximation for the latter function coincides with the mode-coupling
expression derived via the projection operator method. 

To derive a diagrammatic expansion for the time-dependent response function 
we have neglected contributions to the vertices from higher-order terms in
the cluster expansion. It should be noticed that in spite of this fact we
obtained the same mode-coupling expression as the one derived using 
the projection operator method. This suggests that the diagrammatic series
(\ref{Gseries}) contains all the dynamical events that
result in the standard mode-coupling approximation. It would be interesting
to use series (\ref{Gseries}) as a starting point for the development
of theories that go beyond the standard mode-coupling approximation,
\textit{i.e.} to include at least some classes of the diagrams 
that are neglected in the one-loop re-summation. Also, 
it would be interesting to investigate diagrammatic interpretation
of so-called generalized mode-coupling theories \cite{gMCT}. 
Finally, the formalism presented here could be used to derive an approximate
theory for the time-dependence of various four-point correlation functions. Such
functions have been extensively studied in the last decade \cite{four}. 
They provide quantitative information about so-called dynamic heterogeneity
or, more precisely, about correlations of dynamics of different particles.

One of the consequences of neglecting the contributions to the vertices 
from higher-order terms in the cluster expansion is that the approximate 
equations of motion (\ref{notsoexacthierarchy2}-\ref{notsoexacthierarchy2a})
do not reproduce the exact short-time behavior of the density correlation
function. In addition, as noted by Andersen \cite{H3}, 
on physical grounds one would expect
that in the mode-coupling formula (\ref{mmct}) the vertices are replaced by
matrix elements of a binary collision operator. We plan to rectify these 
two drawbacks of the present approach in future work.

The advantage of the present approach is that 
it leads to a relatively simple diagrammatic
series. Thus, it should be possible to derive a field  theoretical representation
of this series. We note, however, that our diagrammatic series is different from series
expansions that have been derived from various field theoretical 
approaches \cite{field1,field2,field3,RC}. First,  
our series involves one dynamical function whereas field-theoretical expansions
are typically formulated in terms of two functions, a correlation function and
a response function (that is different from the response function used in our
formalism). In addition, series (\ref{Gseries}) 
involves both three- and four-leg vertices whereas series expansions resulting from 
field theoretical approaches typically involve only three-leg vertices. Finally,
in our series the renormalization of bare interactions occurs naturally. In
field theoretical approaches one either carries bare interactions throughout or
one has to start from a phenomenological formulation
of dynamics that involves the direct correlation function. 

\section*{Acknowledgments}
I thank Hans Andersen for an inspiring discussion and 
gratefully acknowledge the support of NSF Grant No.~CHE 0517709.

\end{document}